\documentclass{article}
\usepackage{times}
\usepackage{epsf,amsmath}
\usepackage{amssymb}
\usepackage{array,tabularx}

\begin{document}

\title{\Large Collapse of rotating stars in the Universe and the cosmic gamma ray bursts}
\date{A.I. Bogomazov
$^{1}$\thanks{E-mail: a78b@yandex.ru}, V.M. Lipunov
$^{1}$\footnotemark[1]\thanks{E-mail:
lipunov@xray.sai.msu.ru}, A.V. Tutukov$^2$\footnotemark[2]\thanks{E-mail: atutukov@inasan.rssi.ru}\\
{\small $^{1}$ Sternberg astronomical institute, 119992, Universitetskij prospect, 13, Moscow, Russia \\
$^{2}$ Institute for astronomy,  119017, 48, Pyatnitskaya street,
48, Moscow, Russia} }

\maketitle \rm

\label{firstpage}

\begin{abstract}
We analyze here late evolutionary stages of massive ($M_0\gtrsim
8M_{\odot}$) close binary stars. Our purposes are to study
possible mechanisms of gamma ray bursts (GRBs) origin. We suppose
in this paper that GRB phenomenon require formation of massive
($\sim 1 M_{\odot}$) compact ($R~\lesssim~10$ km) accretion disks
around Kerr black holes and neutron stars. Such Kerr black holes
are products of collapse of Wolf-Rayet stars in extremely close
binaries and merging of neutron stars with black holes and neutron
stars with neutron stars in close binary systems. Required
accretion disks also can be formed around neutron stars which were
formed during collapse of accreting oxygen-neon white dwarfs. We
have estimated frequencies of events which lead to a rotational
collapse concerned with formation of rapidly rotating relativistic
objects in the Galaxy. We made our calculations using the
"Scenario Machine".
\end{abstract}


\section{Introduction}

Investigation of the gamma ray bursts (GRBs) physics is one of the
most actual astrophysical problems during last decades and it's
popularity permanently grow. Some years ago frequency of
appearance of the articles on GRB physics had exceeded the
frequency of the recorded GRBs which equals to approximately one
flash per day for the detectors with threshold level $\sim
10^{-7}$ erg cm$^{-2}$ sec$^{-1}$ in the range 30 KeV -- 100 MeV
\cite{batse, postnov1999}. About thirty year have elapsed since
pioneering works about observed GRBs
\cite{klebesadel1973a,usov1975,mazets1986}, but reliable
observational limits on the main parameters of the events have
appeared only during recent years and two conceptions which
probably present two main types of GRBs were accepted. These types
are short (SGRBs) which duration is lower than 2 seconds and long
(LGRBs) which duration is $\sim 2 - 200$ seconds
\cite{kouveliotou1993, janiuk2006a}.

GRBs were supposed to have cosmological nature by Usov and
Chibisov \cite{usov1975}. Now numerous GRBs positions were
identified with positions of galaxies and some supernovae type I
b,c since 1997 \cite{galama1998a}. So, cosmological nature of the
most of the GRBs had become evident
\cite{usov1975,paczynski1986,paczynski1991,lipunov1995a,postnov1999,firmani2004,janiuk2006a}.
It can be illustrated by the isotropy of the GRBs distribution in
the sky and by the deficiency of faint flashes in comparison with
isotropic distribution in the Euclidean space
\cite{lipunov1995a,stern2001,stern2002,firmani2004}. These facts
enable researchers to realize cosmological distances to the most
GRBs and to reconstruct some cosmological parameters, also they
put problem of reconstruction of the star formation history in the
Universe \cite{tutukov2003, firmani2004}. Direct identification of
red shifts of some GRBs (until $z\sim 6$) \cite{haislip2005}
supports this possibility. Identification of the host galaxies
enables to estimate distances to some GRBs and their total energy
in the assumption of spherically symmetric radiation -- $0.1
M_{\odot} c^2$ \cite{postnov1999}, and frequency of the GRBs in a
galaxy with mass like the mass of the Milky Way $\sim
10^{-6}-10^{-7}$ yr$^{-1}$ \cite{mao1992,paczynski1998}.
Assumption about collimation of the gamma radiation in the narrow
space angle ($\sim 0.1$ steradian)
\cite{paczynski1998,matsubayashi2005} allows to get the last
estimations up to the energy $\sim 10^{51}-10^{52}$ erg and
frequency $\sim 10^{-4}-10^{-5}$ yr$^{-1}$. Note that these
numbers are still uncertain by a factor of $\sim 10$. Also it is
important to note that GRBs frequency is less than frequency of
the known supernovae at least several tens times. This
circumstance directed investigators to the most relic, but
powerful events in stellar life. It is important to say that
energetics of SGRBs in X-ray and gamma ray ranges is almost one
hundred times less than energetic of LGRBs \cite{fox2006a}. This
is direct indication that frequency of SGRBs per galaxy can be
higher than frequency of LGRBs. Additional indication of
relatively higher frequency of SGRBs is relatively small distance
to typical identified SGRB ($z\approx 0.2$) with comparison to the
distance to typical GRB (z=2) \cite{kulkarni2005}).

Millisecond variability of observed flux of GRBs is evidence of a
small volume of the main energy-release region, it's size does not
exceed $\sim 10^8$ cm. Three sorts of astrophysical objects with
such dimensions are known: degenerated dwarfs (DD, or white
dwarfs, WD), neutron stars (NS) and black holes (BH). High energy
release $\sim 10^{51}-\sim10^{52}$ erg is typical for NSs and BHs.
It provides for conditions upon ideas how to explain GRB
mechanism. Concept of merge of two NSs under influence of
gravitational waves was advanced to depict SGRBs. Roche lobe
overflow by one of the compact stars leads to it's decay in
dynamical time scale $10^{-4}-10^{-3}$ sec. \cite{tutukov1979}. It
allows to conclude that energy release during such event is enough
to produce a SGRB and to make a relation between a SGRB and a
NS+NS merging \cite{blinnikov1984}. NS also can be a component in
a close binary including BH. List of the four known close binary
systems consisting of neutron stars and radio-pulsars in the
Galaxy which are able to merge in timescale shorter than Hubble
time was compiled \cite{ihm2005}. It includes such systems as
B1913+16 with orbital period about eight hours. According to the
scenario program such pairs also have to merge with frequency
equals to frequency of merging of the close binary systems $1\cdot
10^{-4}$ yr$^{-1}$ \cite{lipun1987a,tutukov1993,tutukov2002}. In
the present article we consider both variants of SGRB formation to
estimate their frequencies. It is necessary to note that in spite
of formal high "delay" of merging after the moment of the
formation of the system (NS+NS, NS+BH) considerable part of
merging happen in first 1-2 Gyr. Although short GRBs can be found
also in old elliptical galaxies \cite{lipun1999a, gorosabel2006a},
most of them have to be related to galaxies with active current
star formation \cite{hjorth2003a}. In fact one short GRB was found
in the galaxy with active star formation \cite{prochaska2006a}.
Observation of the short GRB 050709 in optics allows to exactly
exclude even a faint SN Ic in the same place \cite{hjorth2005b}.
But it is important that in elliptical galaxies there are only
SGRBs
\cite{gal-yam2005,hjorth2005a,lee2005,prochaska2006a,bloom2006a}.
It is evidence in favour of the model of NS+NS or NS+BH mergers.

LGRBs are concerned with collapses of rapidly rotating nuclei of
massive pre-supernova which produce Kerr BH \cite{paczynski1998}.
Collapse of a fast rotating star as mechanism of supernova
explosion was suggested by \cite{biskogan}. Numerical gas-dynamic
model of such event was constructed by \cite{umeda2005}. There are
two possible causes of fast pre-supernova nucleus rotation:
acceleration of the star nucleus rotation by it's grip with
angular momentum conservation \cite{tutukov1969, maeder2000} or
companion presence near pre-supernova helium Wolf-Rayet star (WR)
which is one of components in a close binary
\cite{iben1985,woosley1993a,paczynski1998,postnov2001,tutukov2003,tutukov2004}.
The last version seems to be preferable due to very probable
significant slowing down of rotation of cores of single massive
stars during their evolution \cite{tutukov2003b, meynet2005}.
Orbital period of the system consisting of pre-supernova type I
b,c (WR star) and another component has to be shorter than 1-3
days to produce Kerr black hole. At least one binary including
WR-star (progenitor of GRB) is known at present time: Cyg X-3
\cite{tutukov2003}. There are also some known binary systems
including post-Kerr BH, for instance, V 518 Per
\cite{tutukov2003}. Observational data allow to make relation
between LGRBs and explosions of SN I b,c which mean the end of
WR-star evolution \cite{eldridge2006a,folatelli2006a} in galaxies
with active current star formation
\cite{gorosabel2006a,sollerman2005,bosnjak2006a}. A companion of a
WR-star (a progenitor of a Kerr BH) can be a main sequence star, a
BH or a NS. We investigate here such systems to estimate frequency
of formation of Kerr black holes in them. Note that we treat a
black hole as Kerr BH in this paper if Kerr parameter

\begin{equation}
\label{kerr} a=\frac{I\Omega}{GM_{BH}^2/c}\ge 1,
\end{equation}

\noindent where $I$ is the moment of inertia of the black hole,
$\Omega$ is its angular velocity, $M_{BH}$ is the mass of the
black hole.

It is known that even in very close binaries with helium
non-degenerated progenitors which mass is about $2.5 M_{\odot}$ NS
formation does not allow to produce NS with period of rotation
shorter than $\sim 0.05$ sec. according to elementary
estimations\footnote{ For $M_{He}=2.5M_{\odot}$ practically
filling its Roche lobe with radius $R_{He}=0.34_{\odot}$
\cite{tutukov1973a} major semi-axis $a\simeq R_{\odot}$ for
another companion of solar mass. $P_{orb}=10^4
\frac{a^{3/2}}{M_1+M_2}$ sec. ($a=a/R_{\odot}$, $M=M/M_{\odot}$),
radius of the neutron star is $R_{NS}=10^6$ cm., radius of the
iron core of the helium star is $R_{Fe}=3\cdot 10^8$ cm
\cite{kippenhahn1990}.  $P_{NS}=10^4
\frac{a^{3/2}}{M_1+M_2}\left(\frac{R_{NS}}{R_{Fe}}\right)^2=10^4\frac{1}{\sqrt{
3.5}}\left(\frac{10^6}{3\cdot 10^8}\right)^2\approx 0.06$ sec.}.
At the same time the ultimate lower limit of the period of a
neutron star is much shorter and equals to $\sim 0.001$ sec. It
excludes such NSs from list of progenitors of the GRBs.
Nevertheless probably there is another channel of formation of a
fast rotating single NS due to merging of close binary degenerated
dwarfs. And at least one of the components in the case has to be
oxygen-neon (ONe) white dwarf. Oxygen burning in it does not
produce enough energy to destroy compact dwarf and finally leads
to collapse and formation of a neutron star  \cite{miyaji1980a,
nomoto1980a}. Initial mass of a star in a close binary system has
to be $\sim 8-10 M_{\odot}$ to form oxygen-neon WD
\cite{iben1985}. Collapse of such dwarf during merging of the
components of the close pair will guarantee NS formation with over
critical rotation that leads to formation of a LGRB according to
our model. In general this model is similar to the model of
collapse of a fast rotating WR-star and it is probably possible to
use this model to explain "long" flashes of gamma-rays.

At the same time compact accretion disk can be formed near young
NS and super-critical accretion onto this NS can leads to a
formation of a transient source of high energy photons
\cite{tutukov2004}. A close companion of ONe WD has to be helium,
carbon-oxygen or ONe dwarf. And although observational evidences
of such merge stay not totally clear we have included these events
in our analysis of frequencies. There are some known observable
analogues of close binary degenerated dwarfs which are necessary
to realize such scenario. They were found in the last years during
search for progenitors of supernova stars type Ia. Most of known
close binary degenerated dwarfs have common mass lower
Chandrasekhar limit, but three of them have masses higher than
this limit and only one of them can merge during period of time
$\sim 10^{10}$ years \cite{napiwotzki2004}. This circumstance and
scenario estimations give us basis to hope that it is possible to
produce binary systems consisting of ONe WD and degenerated
companion with common mass higher than Chandrasekhar limit which
also are able to merge under gravitational waves influence during
Hubble time. This is third possible scenario explaining
observation of "long" flashes of gamma rays.

It is necessary to note that although the fact of possible
formation of NS during ONe WD collapse was mentioned
\cite{ruderman2000} more than once, but the picture of the
explosion of accreting matter ONe white dwarf with NS formation is
not known in details yet. Observational manifestations of
degenerated ONe dwarfs with comparable masses have not been
detailed analyzed in spite of relatively high possible frequency
of such events. According to scenario model it is $\sim 0.01$ per
year in the Galaxy \cite{tutukov2002}. But the most part of them
are helium WDs. Duration of the phase of the destruction of a
dwarf is only some seconds. After NS and massive disk formation
the disk will evolve in dissipative time scale depending on the
disk thickness and this time will be $10^2$ -- $10^4$ times longer
than Keplerian time of the disk \cite{shakura,tutukov2004}. That
is, process of destruction of degenerated companion can manifest
as powerful X-ray flash with duration of some hours or days.
Non-recurrent events of such type are actually recorded
\cite{arefiev2003, sguera2005}. It is evidently that observed
X-ray bursts present a heterogeneous group \cite{smith2006a} and
it is a problem for the distant future to pick out among them
events of dynamical disruption of degenerated components in binary
stars. It is important for investigations of GRB models to discuss
how we can explain production of short flashes with magnitude
comparable with SN explosions in close binary systems. Current
classification of GRBs based on their duration is probably not
exhaustive. Investigations of spectra of GRBs allow, for example,
to outline indications of third subgroup of the phenomenon; it has
duration 2-10 seconds and relatively soft gamma-spectra. It is
obtained that optical afterglows during first $\sim 100$ days have
bimodal behaviour \cite{liang2006a}. This result is based on
investigation of about forty light curves of optical "echoes" of
bright GRBs \cite{liang2006a}. Subgroup of short GRBs is also
heterogeneous \cite{tanvir2005}. It is possible that following
detailed study of these phenomena will allow to determine new
sorts of GRBs which will differ in models of progenitors, in
parameters of the models or jet axis orientation relative to
observer.

Formation of young neutron star due to ONe dwarf collapse with
mass higher than Chandrasekhar limit seems to be interesting not
only as potential mechanism of GRB formation, but also as possible
way of magnetars -- X-ray pulsars formation \cite{lipun1985a}.
Only one of them -- XTE J1810-197 was revealed as transient
radio-pulsar. Magnetars are neutron stars with magnetic fields
$\sim 10^{15}$ Gs and rotational periods 5-12 sec. and current
estimation of their birth frequency $\sim 10^{-3}$ per year
\cite{thompson1993,kouveliotou1999,thompson2001,popov2006a}. It is
well known that magnetic fields of some degenerated dwarfs reach
to $\sim 10^8$-$10^9$ Gs \cite{vanlandingham2005}. During collapse
of such dwarf with radius $\sim 10^9$ cm into NS with radius
$10^6$ cm trapped magnetic field will be intensified by a factor
of million, i.e. will reach to $\sim 10^{14}$-$10^{15}$ Gs
(observed value, see \cite{woods2004a} for details). Slow rotation
of magnetars is probably consequence of their very strong field
which rapidly increases their rotational periods from initial,
which are probably very short, up to observed value
\cite{beskin2005}. To accelerate ONe dwarf rotation to the
quantity enough to form Keplerian disk around young neutron star
after dwarf collapse, i.e. GRB, it is enough according to simple
estimation to accrete a part of matter of the Keplerian disk with
mass $\Delta M/M_{\odot}\simeq (R_{NS}/R_{ONe})^{\frac{1}{2}}\sim
0.03 M_{\odot}$, where $R_{NS}$ and $R_{ONe}$ -- radii of NS and
degenerated ONe dwarf correspondingly.

\section{Population synthesis}

We use the "Scenario Machine" to estimate frequencies of the
described above events which can lead to GRB formation. For every
set of the initial parameters we have conducted population
synthesis of $10^6$ binary systems.

Since the "Scenario machine" working principles were described
more than once, in present work we will limit ourself only with
mentioning of the most important parameters which influence on the
results of numerical modelling binaries under investigation.
Detailed description of the "Scenario machine" may be found in the
book written by \cite{lipunov1996}.

\subsection{Initial mass ratio distribution}

Population synthesis was conducted for two types of the initial
distribution of the mass ratio of the components in a binary
system $f(q)=q^{\alpha_q}$: flat ${\alpha_q}=0$ (the most probable
value, see \cite{krajcheva1981} for details) and quadratic
${\alpha_q}=2$ (see \cite{lipunov1995b} for details). As the
coefficient of the mass ratio in a binary system $q=m_2/m_1$ we
assume the ratio of the mass of the secondary companion $m_2$ to
the mass of the primary companion $m_1$ of the system ($m_1>m_2$).
Note that for low mass systems ($m_1 \le 10 M_{\odot}$) we take
$\alpha_q=0$ in all scenarios.

\subsection{'Kick' during supernova explosion}

It was supposed in our calculations that neutron star or black
hole during supernova explosion is able to get additional "kick"
$v$, it's velocity distributed by Maxwell function:

\begin{equation}
\label{Maxwell} f(v)\sim \frac{v^2}{v^2_0}e^{-\frac{v^2}{v_0^2}},
\end{equation}

\noindent it's direction is equiprobable. But the quantity of
dispersion $v_0$ of the remnant is one of the crucial parameters
for estimations of the frequencies. We should to say that the
results of population synthesis are highly dependent on the
quantity of the parameter $v_0$. Increasing $v_0$ higher than
orbital velocity in close binary systems $\sim 100$ km s$^{-1}$
leads to sharp reduction of number of systems including
relativistic companion.

Let us suppose that absolute value of kick velocity during BH
formation depends on the part of the mass loss by an exploding
star during supernova explosion. In our calculations we assumed
that during supernova explosion a star lose a half of it's mass
(see \cite{bogomazov2005} for details). If we assumed
characteristic velocity of the neutron star kick $v_0$, the
quantity of the parameter $v_0$ for the case of formation of the
black holes is defined as $v_{0}^{bh}=0.5v_0$ in present work.

\subsection{Stellar wind}

Mass loss by an optical star during its evolution is still poorly
known at present time. Although it is possible to significantly
reduce uncertainties (see, for instance, \cite{bogomazov2005}),
there are no ample reason to choose single scenario of mass
outflow by stellar wind as a standard, so population synthesis was
done using two different scenarios of mass loss rate by a
non-degenerated star. Let us call them A and C.

Stellar wind significantly influences on evolution of the massive
stars which core's collapse can lead to a GRB formation. In our
conception the phenomenon of GRB can happen only in a close binary
system. It's components either merge with Kerr black hole
formation or rotational collapse of WR-star (into Kerr black hole)
or white dwarfs (into neutron star) is enabled due to orbital
motion. Optical star mass loss greatly influences on the major
semi-axis of binary system and is able to significantly change
number of close binary systems which can produce GRB.

In this work we use quasi-conservative mass transfer (see
\cite{lipunov1996,heuvel1994} for details). In this case we
calculate the major semi-axis of the system using formula

\begin{equation}
\label{semiaxis}
\frac{a_f}{a_i}=\left(\frac{q_f}{q_i}\right)^3\left(\frac{1+q_i}{1+q_f}\right)\left(\frac{1+\frac{\beta}{q_f}}{1+\frac{\beta}{q_i}}\right)
\end{equation}

\noindent where $a_f$ is the final major semi-axis, $a_i$ is the
initial major semi-axis. In this equation $q_f$ and $q_i$ are
final and initial values of $q=M_{accr}/M_{donor}$, here
$M_{accr}$ is the mass of the accreting star and $M_{donor}$ is
the mass of the donor star. Parameter
$\beta\equiv-(M_{accr}^i-M_{accr}^f)/(M_{donor}^i-M_{donor}^f)$ we
calculate in this case as minimal value between $\beta=1$ and the
ratio $\beta=T_{KH}(donor)/T_{KH}(accr)$, where $T_{KH}(donor)$ is
the Kelvin-Helmholtz time for the donor, $T_{KH}(accr)$ is the
same for the accretor.

Scenario A has a weak stellar wind. The mass loss rate $\dot M$
during the main sequence (MS) stage \cite{jager} is

\begin{equation}
\dot M\sim L/V_{\infty}, \label{mlossa1}
\end{equation}

\noindent where $L$ is the luminosity of the star and $V_{\infty}$
is the wind velocity at infinity.

For giants we take a maximum between (\ref{mlossa1}) and the
result obtained by \cite{lamers}:

\begin{equation}
\dot M\sim L^{1.42}R^{0.61}/M^{0.99}, \label{mlossa2}
\end{equation}

\noindent where $R$ is the stellar radius, $M$ is its mass.

For red supergiants we take a maximum between (\ref{mlossa1}) and
Reimers's formula \cite{kudritzki}:

\begin{equation}
\dot M\sim LR/M, \label{mlossa3}
\end{equation}

The mass change $\Delta M$ in wind type A during one stage (except
WR-stars) is no more than $0.1(M-M_{core})$, where $M$ is the mass
of the star at the beginning of a stage and $M_{core}$ is its core
mass. Mass loss during the Wolf-Rayet (WR) star stage is
parametrized as $0.3\cdot M_{WR}$, where $M_{WR}$ is the maximum
star mass during this stage. For calculations of stellar wind type
A we used the core masses obtained by \cite{varshavskii} and
\cite{iben1985,iben2}.

In scenario C the stellar evolution model is based on the results
of \cite{vanbeveren}, which reproduce most accurately the observed
galactic WR star distributions and stellar wind mass loss in
massive stars. Calculations of mass loss by a star were conducted
if we used the formula

\begin{equation}
\Delta M=(M-M_{core}), \label{m_A}
\end{equation}

\noindent where $M_{core}$ is the stellar core mass
($\ref{m_core_C}\alpha\div \ref{m_core_C}\epsilon$). If the
maximum mass of a star (usually it is initial mass of a star, but
mass transfer in binary system is able to increase its mass over
the initial value) $M_{max}>15M_{\odot}$ the mass of the core in
the main sequence stage is determined using
($\ref{m_core_C}\alpha$), and in giant and in supergiant stages
using ($\ref{m_core_C}\beta$). In the Wolf-Rayet star stage, if
$M_{WR}<2.5M_{\odot}$ and $M_{max}\le 20M_{\odot}$ it is described
using ($\ref{m_core_C}\gamma$), if $M_{WR}\ge 2.5M_{\odot}$ and
$M_{max}\le 20M_{\odot}$ as ($\ref{m_core_C}\delta$), if
$M_{max}>20M_{\odot}$ using ($\ref{m_core_C}\epsilon$).

\begin{equation}
M_{core} = \left\{
\begin{array}{l}
1.62M_{max}^{0.83}, \qquad\qquad\qquad\qquad\qquad\;\;\,\, (\alpha) \\
10^{-3.051+4.21\lg M_{max} -0.93(\lg M_{max})^2}, \; (\beta) \\
0.83M_{WR}^{0.36}, \qquad\qquad\qquad\qquad\qquad\;\;\;\, (\gamma) \\
1.3+0.65(M_{WR}-2.4), \qquad\qquad\;\;\;\;\,\, (\delta) \\
3.03M_{max}^{0.342}, \qquad\qquad\qquad\qquad\qquad\;\;\, (\epsilon) \\
\end{array}\right.
\label{m_core_C}
\end{equation}

Scenario C has high mass loss during the WR stage, it may reach
$50\%$ of a star mass or more here. Mass loss in other stages (MS,
giant, supergiant) for stars with masses higher than $15M_{\odot}$
(for less massive stars this scenario is equivalent to a type A
wind) may reach $\approx 30\%$ of the mass of a star. Total mass
loss $\Delta M$ during all stages always is larger than in
scenario A.

\subsection{Common envelope stage efficiency}

An effective spiral-in of the binary components occurs during the
common envelope (CE) stage. The effectiveness of the CE-stage is
described by the parameter $\alpha_{CE}=\Delta E_b/\Delta
E_{orb}$, where $\Delta E_b=E_{grav}-E_{thermal}$ is the binding
energy of the ejected envelope matter and $\Delta E_{orb}$ is the
drop in the orbital energy of the system during spiral-in
\cite{heuvel1994}.

\begin{equation}
\alpha_{CE} \left( \frac{GM_a M_c}{2a_f} - \frac{GM_a M_d}{2a_i}
\right)=\frac{GM_d(M_d - M_c)}{R_d},
\end{equation}

\noindent where $M_c$ is the mass of the core of the mass-losing
star of initial mass $M_d$ and radius $R_d$ (which is simply a
function of the initial separation $a_i$ and the initial mass
ratio $M_a/M_d$, where $M_a$ is the mass of the accreting star).

\subsection{Restrictions on key parameters of binary evolution}

Previous estimations of possible value area of parameters of
binary stars evolution were made in papers
\cite{lipunov1996b,lipunov1997}. But during these years some new
results have appeared and we made some additional calculations.

Newest observational estimations of neutron star kick during
supernova explosion is \cite{hobbs2005}. The authors concluded
that characteristic kick velocity $v_0=265$ km s$^{-1}$.

\cite{dewi2000} suggested to make more correct estimations for
common envelope stage efficiency. They argued to make correction
of gravitational energy taking into account the fact that matter
of the stars concentrates to their center $\frac{GM_d(M_d -
M_c)}{R_d\lambda}$. But in their calculations they supposed common
envelope stage efficiency $\mu_{CE}$ to be equal to 1. In general,
we don't know this parameter exactly. Our coefficient
$\alpha_{CE}$ is the production of the common envelope efficiency
$\mu_{CE}$ and the parameter $\lambda$. So we use estimations of
$\alpha_{CE}$ suggested by \cite{lipunov1996b}.

We would like to note one important circumstance. Poorly known
evolution parameters, such as $v_0$, $\alpha_q$, stellar wind,
etc. are intrinsic parameters of population synthesis. In future
they can be defined more exactly or their physical context can be
changed: distribution of kick velocity might not be Maxwell,
complicated hydrodynamics of common envelope probably can not be
described in terms of $a_{ce}$ and $\lambda$, distribution $f(q)$
might not be power law. So, we have only one way to move ahead in
investigations, what is to compare predictions of our models with
observational data.

\begin{figure}
\vspace{0pt}\epsfbox{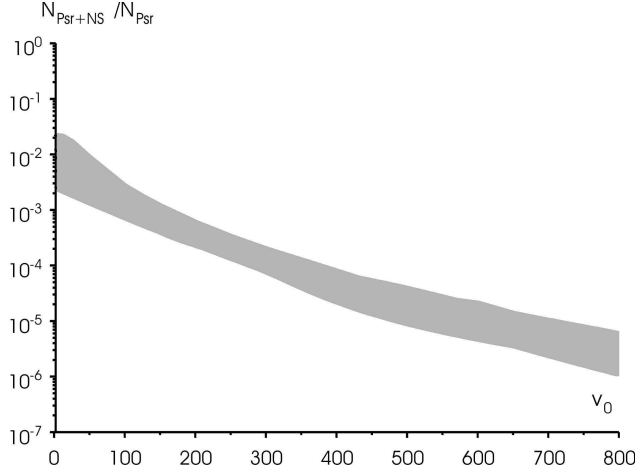}\caption{ This figure shows how
ratio $\frac{N_{NS+Psr}}{N_{Psr}}$ (here $N_{NS+Psr}$ is the
calculated number of binary neutron stars with radio pulsars and
$N_{Psr}$ is the calculated number of all radio pulsars) depends
on two parameters: kick velocity $v_0$ and common envelope stage
efficiency $\alpha_{CE}$. "Width" of the filled area depicts
various values of $\alpha_{CE}$ in the range between 0.2 and 1.0.
}
\end{figure}

\begin{figure}
\vspace{0pt}\epsfbox{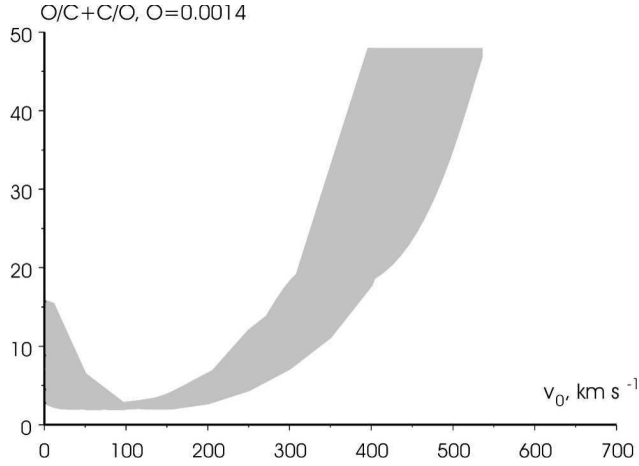}\caption{ This figure shows OCCO
criterion \cite{lipunov1996} for ratio
$\frac{N_{NS+Psr}}{N_{Psr}}$. $v_0$ is NS's characteristic kick
velocity. "Width" of the filled area depicts various values of
$\alpha_{CE}$ in the range between 0.2 and 1.0. Note that
observation value of this ratio is $\sim 0.001$. }
\end{figure}

\begin{figure}
\vspace{0pt}\epsfbox{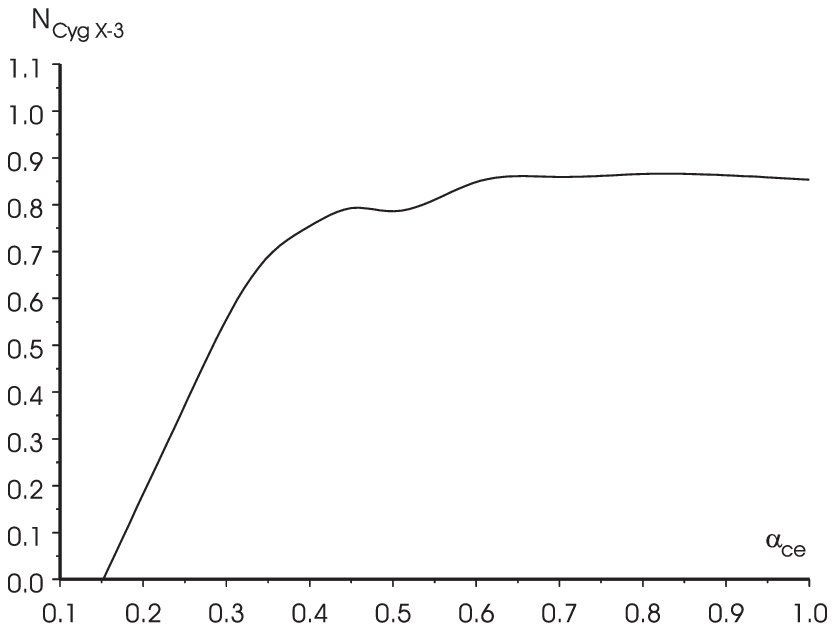}\caption{ Model number of Cyg X-3
type systems in the Galaxy as the function of the common envelope
stage efficiency. }
\end{figure}

We suggest here two tests for our model: comparisons between
calculated number of systems of Cyg X-3 type and ratio
$\frac{N_{NS+Psr}}{N_{Psr}}$ (number of radio pulsar in binary
systems with a neutron star divided by total number of radio
pulsars, binary and single). To avoid differences between young
and recycled radio pulsar we take only young pulsars. Note that
current observational estimation of value of this ratio is $\sim
0.001$ (more exactly, catalogue duplicity rate, \cite{atnf}):
there are two known young binary radio pulsars with a neutron star
companion (J2305+4707 and J0737-3039) and $> 1500$ known radio
pulsars. As Cyg X-3 type system we take here a black hole with
WR-star, mass of WR-star is $>7 M_{\odot}$ and orbital period of
the pair is less than 10 hours.

In the Figure 1 we show how ratio $\frac{N_{NS+Psr}}{N_{Psr}}$
(here $N_{NS+Psr}$ is the calculated number of binary neutron
stars with radio pulsars and $N_{Psr}$ is the calculated number of
all radio pulsars) depends on two parameters: kick velocity $v_0$
and common envelope stage efficiency $\alpha_{CE}$. "Width" of the
filled area depicts models with $\alpha_{CE}$ in the range between
0.2 and 1.0. We used stellar wind type A, $\alpha_q=0$ for these
calculations.

In the Figure 2 we show OCCO criterion \cite{lipunov1996} for
ratio $\frac{N_{NS+Psr}}{N_{Psr}}$: $\frac{O}{C}+\frac{C}{O}$,
where $O$ is the observed number of the quantity and $C$ is the
calculated number of the quantity. If $O=C$ this value is equal to
2. As one can see from the Figure 2 high kick velocity $v_0>200$
km s$^{-1}$ leads to lack of binary radio pulsars with neutron
stars.

In the Figure 3 we present estimated with our model number of Cyg
X-3 systems. As we can see from this figure, they are practically
exclude values of $\alpha_{CE}< 0.3$.

\section{Results and discussions}

Frequencies of events which can produce a GRB calculated using
"scenario machine" are presented the tables \ref{table0},
\ref{table1} and \ref{table2}. All shown frequencies are
normalized to a galaxy with mass and star formation rate equal to
the mass and the current star formation rate in the Milky Way.
This suggestion is valid, because even in case of mergers most of
the events (merging and collapsing) must happen during first
billion years \cite{tutukov1993, lipunov1995a} from formation of
the appropriate systems. It is important to note that the minimal
mass of the star which evolution remnant is the black hole assumed
in the present work to be equal to 25 masses of the Sun
\cite{tutukov2003}.

In the table \ref{table0} we have presented frequencies of events
which are able to produce a GRB. In the first place we put merging
of oxygen-neon white dwarfs with oxygen-neon, carbon-oxygen and
helium white dwarfs and accretion induced collapses (AIC) of
oxygen-neon white dwarfs in systems with optical companion. Also
we showed in the table \ref{table0} frequencies of merging neutron
stars with neutron stars and neutron stars with black holes. So,
in the table \ref{table0} we have showed frequencies calculated
using the next scenario parameters: stellar wind A,
$\alpha_{ce}=0.5$, $\alpha_q=0$, characteristic kick velocity
$v_0=0$ for neutron stars and black holes. In the table
\ref{table1} frequencies of rotational collapses of Wolf-Rayet
stars in close binary systems are presented. Since the critical
period of rotation of WR-star is not reliably fixed yet we have
made our calculations using two values of critical orbital period
$P_{crit}$, where $P_{crit}$ is the maximum orbital period of a
close binary system in which GRB is able to be formed.
Calculations of rotational collapses of WR-stars we take into
account three types of close binary systems: consisting of a black
hole and a Wolf-Rayet star (BH+WR), WR-star and main sequence star
(WR+MS), WR-star and non-degenerated star in Roche lobe overflow
stage (WR+Rlo). It is necessary to note that in case of stellar
wind C minimal period of a binary star including WR-star just
before collapse becomes to about five days; it is much higher than
estimation of the maximum period of binary in which GRB is able to
be formed.

\begin{table}
\caption{Frequencies of events in the Galaxy which are able to
produce a GRB. Stellar wind A, $\alpha_{ce}=0.5$,
$\alpha_q=0$.}\label{table0}
\begin{minipage}{75mm}
\centering
\begin{tabular}{@{}cc@{}}
\hline
 Event & Frequency, yr$^{-1}$  \\
\hline
\multicolumn{2}{c}{ White dwarfs }\\
\hline
 ONe+CO         & $1.8 \cdot 10^{-3}$  \\
 ONe+He         & $1.7 \cdot 10^{-5}$  \\
 ONe+ONe         & $4.9 \cdot 10^{-4}$ \\
 ONe AIC          & $8 \cdot 10^{-5}$  \\
\hline
\multicolumn{2}{c}{ Hyper-nova model} \\
\hline
\multicolumn{2}{c}{ $P_{crit}=1$ day}\\
\hline
 BH+WR        & $3.7\cdot 10^{-6}$  \\
 WR+MS        & $3.4\cdot 10^{-5}$  \\
 WR+Rlo       & $1.1\cdot 10^{-5}$  \\
\hline
\multicolumn{2}{c}{ $P_{crit}=3$ days} \\
\hline
 BH+WR    & $9\cdot 10^{-6}$     \\
 WR+MS      & $2.6\cdot 10^{-4}$   \\
 WR+Rlo    & $2.9\cdot 10^{-5}$     \\
\hline
\multicolumn{2}{c}{Neutron stars with neutron stars}\\
\multicolumn{2}{c}{and neutron stars with black holes}\\
\multicolumn{2}{c}{mergers, kick velocity $v_0=0$ }\\
\hline
NS+NS & $2.1\cdot 10^{-4}$ \\
NS+BH & $7.4\cdot 10^{-5}$ \\
\hline
\end{tabular}
\begin{minipage}{75mm}\small Comment: $P_{crit}$ is the minimal orbital period of close binary system in which rotational collapse of Wolf-Rayet star is possible to produce GRB. \end{minipage}
\end{minipage}
\end{table}

In the tables \ref{table1} and \ref{table2} one can find how
different scenario parameters influence on the calculated
frequencies of events which we are studying in this work.

In the table \ref{table1} we have presented frequencies of merging
of oxygen-neon white dwarf with oxygen-neon, carbon-oxygen and
helium white dwarfs and accretion induced collapses (AIC) of
oxygen-neon white dwarfs in systems including optical companion
for $\alpha_{CE}=1.0$. Note that in all cases we calculate low
mass ($m_1\le 10M_{\odot}$) systems using stellar wind A and
$\alpha_{q}=0$. Also in the table \ref{table1} we have presented
frequencies of rotational collapses of Wolf-Rayet stars in close
binary systems which were calculated using different scenario
parameters.

\begin{table}
\caption{Frequencies of collapses of Wolf-Rayet stars and ONe
white dwarfs in the Galaxy calculated using different scenario
parameters. Stellar wind A.}\label{table1}
\begin{minipage}{75mm}
\centering
\begin{tabular}{@{}cc@{}}
\hline
 Event & Frequency, yr$^{-1}$  \\
\hline
\multicolumn{2}{c}{White dwarfs, $\alpha_{CE}=1$}\\
\hline
 ONe+CO         & $1.1 \cdot 10^{-3}$  \\
 ONe+He         & $7.2 \cdot 10^{-5}$  \\
 ONe+ONe         & $4.5 \cdot 10^{-4}$ \\
 ONe AIC          & $2.7 \cdot 10^{-5}$  \\
\hline
\multicolumn{2}{c}{ Hyper-nova model }\\
 \hline
\multicolumn{2}{c}{ $\alpha_q=2$, $\alpha_{ce}=0.5$ }\\
\hline
 \multicolumn{2}{c}{ $P_{crit}=1$ day }\\
 \hline
 BH+WR    & $3.2\cdot 10^{-6}$ \\
 WR+MS     & $4.6\cdot 10^{-6}$    \\
 WR+Rlo      & $2.2\cdot 10^{-6}$ \\
 \hline
 \multicolumn{2}{c}{ $P_{crit}=3$ days }\\
 \hline
 BH+WR     & $1.1\cdot 10^{-5}$     \\
 WR+MS  & $8.2\cdot 10^{-5}$     \\
 WR+Rlo   & $1.1\cdot 10^{-5}$     \\
 \hline
\multicolumn{2}{c}{ $\alpha_q=0$, $\alpha_{ce}=1$ }\\
\hline
\multicolumn{2}{c}{ $P_{crit}=1$ day }\\
\hline
 BH+WR        & $2\cdot 10^{-5}$  \\
 WR+MS        & $4.1\cdot 10^{-5}$      \\
 WR+Rlo    & $8.3\cdot 10^{-6}$   \\
\hline
\multicolumn{2}{c}{ $P_{crit}=3$ days }\\
\hline
 BH+WR        & $3.6\cdot 10^{-5}$  \\
 WR+MS        & $2.9\cdot 10^{-4}$      \\
 WR+Rlo    & $2\cdot 10^{-5}$   \\
 \hline
\end{tabular}
\begin{minipage}{75mm}\small \center Comment:
$P_{crit}$ -- the same as in Table \ref{table0}.
\end{minipage}
\end{minipage}
\end{table}

\begin{table}
\caption{Frequencies of merging of neutron stars with neutron
stars and neutron stars with black holes under different
evolutionary scenario parameters.}\label{table2}
\begin{minipage}{75mm}
\centering
\begin{tabular}{@{}ccc@{}}
\hline
 NS+NS, yr$^{-1}$ & NS+BH, yr$^{-1}$ & $v_{0}$, km s$^{-1}$  \\
\hline
\multicolumn{3}{c}{ Stellar wind A, $\alpha_q=0$ }\\
\hline
$7.3\cdot 10^{-5}$ & $6.5\cdot 10^{-5}$ & 50 \\
$2.3\cdot 10^{-5}$ & $3.7\cdot 10^{-5}$ & 100 \\
$7.3\cdot 10^{-6}$ & $1.7\cdot 10^{-5}$ & 200 \\
$2.6\cdot 10^{-6}$ & $9.8\cdot 10^{-6}$ & 300 \\
\hline
\multicolumn{3}{c}{ Stellar wind A, $\alpha_q=2$ }\\
\hline
$4\cdot 10^{-4}$ & $6.1\cdot 10^{-5}$ & 0 \\
$1.4\cdot 10^{-4}$ & $5.2\cdot 10^{-5}$ & 50 \\
$4.7\cdot 10^{-5}$ & $2.8\cdot 10^{-5}$ & 100 \\
$1.7\cdot 10^{-5}$ & $9.4\cdot 10^{-6}$ & 200 \\
$6\cdot 10^{-6}$ & $5\cdot 10^{-6}$ & 300 \\
\hline
\multicolumn{3}{c}{ $v_{0}=0$, $\alpha_q=0$ }\\
\hline
\multicolumn{3}{c}{ Stellar wind C }\\
\hline
 NS+NS, yr$^{-1}$ & NS+BH, yr$^{-1}$ &  $\alpha_{CE}$ \\
\hline
$2\cdot 10^{-4}$ & $2.3\cdot 10^{-5}$ & 0.5 \\
\hline
\multicolumn{3}{c}{ Stellar wind A }\\
\hline
$2.4\cdot 10^{-4}$ & $3.9\cdot 10^{-5}$ & 1.0\\
\hline
\end{tabular}
\end{minipage}
\end{table}

In the table \ref{table2} one can find frequencies of merging of
neutron stars with neutron stars and neutron stars with black
holes if we used various scenario parameters described above.

During analysis of the frequencies showed in the tables
\ref{table0}, \ref{table1} and \ref{table2} it is necessary to
take into account collimation of gamma ray emission in the small
space angle \cite{postnov1999}. It means that only a small part of
the GRBs might be observed on the Earth. This fact allows to
estimate GRB frequency in the Galaxy: $\sim 10^{-4} - 10^{-5}$ per
year \cite{mao1992,postnov1999}. And it is quite possible that
collimation depends on the scenario, so table quantities can not
be directly related to observed frequencies. Comparison of this
quantity with theoretical estimations of events probably forming
GRB flashes (see table \ref{table1}) leads us to a conclusion
about potential adequacy of these frequencies for the explanation
of observable frequency of GRBs. However note that all these
frequencies remain rather uncertain at present time. Analyzing
table \ref{table1} it is important to notice considerable
frequency of ONe degenerated dwarfs merging. For hyper-nova
(WR-star collapse) it is necessary flat to use ($\alpha_q=0$)
initial mass ratio distribution and possible maximum orbital
period limit $\sim 1$ day. Influence of a model of stellar wind on
the frequency of NS+NS or NS+BH merging is inessential, but
neutron star's or black hole's "kick" with magnitude higher than
$\sim 100$ km sec$^{-1}$ has to be declined if observed GRBs are
the products of the phenomena studied in this article
\cite{lipunov2006s}.

We would like also to note that frequencies of WR-stars collapses
have no significant differences in cases of critical orbital
periods $P_{crit}=1$ day and $P_{crit}=3$ days, but if we assume
$P_{crit}\lesssim 0.5$ day all scenarios of GRBs with Wolf-Rayet
stars must be declined because of zero frequency of such events in
frames of our models.

\label{lastpage}


\begin{thebibliography}{99}

\bibitem{arefiev2003} Arefiev V.A., Priedhorsky W.C., Bororzdin K.N., 2003, ApJ, 586, 1238

\bibitem{arnaud2005} Arnaud M., 2005, preprint
(astro-ph/0508159)

\bibitem{atnf} the Australia Telescope National Facility
(ATNF) pulsar catalogue, avialable at:
http://www.atnf.csiro.au/research/pulsar/psrcat/

\bibitem{beskin2005} Beskin V.S., Eliseeva S.A., 2005, Astronomy Letters, 31, 263

\bibitem{biskogan} Bisnovatyi-Kogan G.
S., SvA, 14, 652

\bibitem{blinnikov1984} Blinnikov S.I., Novikov I.D., Perevodchikova T.V., Polnarev A.G., 1984, SvA Lett., 10, 177

\bibitem{bloom2006a} Bloom J. et
al., 2006, ApJ, 638, 354

\bibitem{bogomazov2005} Bogomazov A.I., Abubekerov M.K., Lipunov V.M., 2005, Astronomy Reports, 49, 644

\bibitem{bosnjak2006a}  Bosnjak Z., Celotti A., Ghirlanda G., Della Valle M., Pian E., 2006,
A\&A, 447, 121

\bibitem{conselice2005} Conselice C.J. et al., 2005, ApJ, 633, 29

\bibitem{dewi2000} Dewi J.D.M., Tauris T.M., 2000, A\&A,
360, 1043

\bibitem{eldridge2006a}  Eldridge J.J., Genet F., Daigne F., Mochkovitch
R., 2006, MNRAS, 367, 186

\bibitem{firmani1992} Firmani C., Tutukov A., A\&A, 1992, 264, 37

\bibitem{firmani1994} Firmani C., Tutukov A., A\&A, 1994, 288, 713

\bibitem{firmani2004} Firmani C., Avila-Reese V., Ghisellini G., Tutukov A.V., 2004, ApJ, 611, 1033

\bibitem{folatelli2006a} Folatelli G. et al., 2006,
ApJ, 641, 1039

\bibitem{fox2006a} Fox D. et al.,
2006, Nature, 437, 845

\bibitem{galama1998a} Galama
T.J. et al., 1998, Nature, 395, 670

\bibitem{gal-yam2005} Gal-Yam V., 2005,
preprint (astro-ph/0509891)

\bibitem{ghisellini2005} Ghisellini G., Ghirlanda G., Firmani C., Lazzati D., Avila-Reese V., 2005,
Il Nuovo Cimento C, v. 28, Issue 4, p. 639 (preprint
astro-ph/0504306)

\bibitem{gorosabel2005a} Gorosabel J. et al., 2005,
A\&A, 444, 711

\bibitem{gorosabel2006a} Gorosabel J. et al., 2006,
A\&A, 450, 87

\bibitem{haislip2005} Haislip J., 2005, preprint
(astro-ph/0509660)

\bibitem{halpern2005}  Halpern J.P., Gotthelf E.V., Becker R.H., Helfand D.J., White
R.L., 2005, ApJ, 632, L29

\bibitem{hjorth2003a} Hjorth J. et al., 2003, Nature, 423,
847

\bibitem{hjorth2005a} Hjorth J. et al., 2005, ApJ, 630, L117

\bibitem{hjorth2005b} Hjorth J. et
al., 2005, Nature, 437, 859

\bibitem{hobbs2005} Hobbs G., Lorimer D.R., Lyne A.G., Kramer M., 2005, MNRAS, 360,
974

\bibitem{horvath2004a} Horvath I., Meszaros A., Balazs L.G., Bagoly Z., 2004, Baltic Astron., 13, 217

\bibitem{iben1985} Iben I., Tutukov A., 1985, ApJS, 58, 661

\bibitem{iben2} Iben Icko Jr., Tutukov Alexander V., 1987, ApJ, 313, 727

\bibitem{ihm2005} Ihm C.M., Kalogera V., Belczynski K., 2005, preprint
(astro-ph/0508626)

\bibitem{jager} Jager C., 1980, The Brightest Stars, Reidel,
Dordrecht

\bibitem{janiuk2006a} Janiuk A., Czerny B., Moderski R., Cline D.B., Matthey C., Otwinowski S., 2006,
MNRAS, 365, 874

\bibitem{jorgensen1997a} Jorgensen H.E., Lipunov V.M., Panchenko I.E., Postnov K.A., Prokhorov M.E., 1997, ApJ, 486, 110

\bibitem{kippenhahn1990} Kippenhahn R., Weigert A., 1990, Stellar Structure and Evolution,
Springer-Verlag, p. 366

\bibitem{klebesadel1973a} Klebesadel R.W., Strong I.B., Olson
R.A., 1973, ApJ, 182, L85

\bibitem{kouveliotou1993} Kouveliotou C., Meegan C.A., Fishman G.J., Bhat N.P., Briggs M.S., Koshut T.M., Paciesas W.S., Pendleton G.N., 1993, ApJ, 413,
L101

\bibitem{kouveliotou1999} Kouveliotou C. et al., 1999, ApJ, 510, L115

\bibitem{krajcheva1981}
Krajcheva Z.T., Popova E.I., Tutukov A.V., Yungelson L.R., 1981,
SvA Lett., 7, 269

\bibitem{kudritzki} Kudritzki B.P. and Reimers D., 1978, A\&A,
70, 227

\bibitem{kulkarni2005} Kulkarni S.R., 2005, preprint
(astro-ph/0510256)

\bibitem{kurbatov2005} Kurbatov E.P., Tutukov A.V., Shustov B.M., 2005, Astronomy Reports, 49, 510

\bibitem{lamers} Lamers H. J. G. L. M., 1981, ApJ, Part 1, 245, 593

\bibitem{lee2005} Lee W., Ramirez-Ruiz E., Granot J., 2005, ApJ, 630, L165

\bibitem{liang2006a} Liang E., Zhang B., 2006, ApJ, 638, L67

\bibitem{lipun1983a} Lipunov V.M., 1983, Ap\&SS, 97, 121

\bibitem{lipunov2006s} Lipunov V.M., 2006, Proceedings IAU Symposium No. 230, Cambridge University Press, p. 391.

\bibitem{lipun1985a} Lipunov V.M., Postnov K.A., 1985, A\&A, 144, L13

\bibitem{lipun1987a} Lipunov V.M., Postnov K. A., Prokhorov M.E., 1987, A\&A, 176, L1

\bibitem{lipunov1995a} Lipunov V.M., Postnov K.A., Prokhorov M.E., Panchenko I.E., Jorgensen H.E., 1995, ApJ, 454, 593

\bibitem{lipunov1995b} Lipunov V.M., Nazin S.N., Panchenko I.E., Postnov K.A., Prokhorov M.E., 1995, A\&A, 298, 677

\bibitem{lipunov1996} Lipunov V.M., Postnov K.A., Prokhorov M.E., 1996, ed. Sunyaev R.A., The Scenario Machine: Binary Star Population
Synthesis, Astrophysics and Space Physics Reviews, vol. 9, Harwood
academic publishers

\bibitem{lipunov1996b} Lipunov V.M., Postnov K.A.,
Prokhorov M.E., 1996, A\&A, 310, 489

\bibitem{lipunov1997} Lipunov V.M., Postnov K.A.,
Prokhorov M.E., 1997, MNRAS, 288, 245

\bibitem{maeder2000} Maeder A., Meynet G., 2000, ARA\&A, 38, 143

\bibitem{matsubayashi2005}  Matsubayashi T., Yamazaki R., Yonetoku D., Murakami T., Ebisuzaki T., 2005, preprint
Progress of Theoretical Physics, 114, 983

\bibitem{mao1992} Mao S., Paczynski B., 1992, ApJ, 388,
L45

\bibitem{mazets1986} Mazets E.P., 1986, Proc 19th Cosmic Ray Conf., ed.
F.C. Jonas et al, Washington DC, NASA, p. 415

\bibitem{meynet2005} Meynet G., Maeder A., 2005, A\&A, 429, 581

\bibitem{miyaji1980a} Miyaji S., Nomoto K., Yokoi K., Sugimoto
D., 1980, Asrton. Soc. of Japan Publications (PASJ), 32, 303

\bibitem{napiwotzki2004} Napiwotzki R. et al., 2004, Rev. Mech. AA, v. 20, p. 113

\bibitem{nomoto1980a}
Nomoto K., 1980, in: Type I supernovae, Proceedings of the Texas
Workshop, Austin, p. 164.

\bibitem{batse}
Paciesas W.S. et al., 1999, ApJS, 122, 465

\bibitem{paczynski1986} Paczynski B., 1986, ApJ, 308, L43

\bibitem{paczynski1991} Paczynski B., 1991, Acta Astron., 41, 257

\bibitem{paczynski1998} Paczynski B., 1998, ApJ, 494, L45

\bibitem{panter2004} Panter B., Heavens A., Jimenez R., 2004, MNRAS, 355, 764

\bibitem{panagia2005} Panagia N., Fall S.M., Mobasher B., Dickinson M., Ferguson H.C., Giavalisco M., Stern D., Wiklind T., 2005, ApJ, 633, L1

\bibitem{lipun1999a} Panchenko I.E., Lipunov V.M., Postnov K.A., Prokhorov M., 1999, A\&AS, 138, 517

\bibitem{pgonz2005} Perez-Gonzalez P.G. et
al., 2005, ApJ, 630, 82

\bibitem{postnov1999} Postnov K.A., Physics-Uspekhi, 1999, 42, 469

\bibitem{postnov2001} Postnov K.A., Cherepashchuk A.M., 2001, Astronomy Reports, 45, 517

\bibitem{prochaska2006a} Prochaska J.X. et
al., 2006, ApJ, 642, 989

\bibitem{popov2006a} Popov S.B., Prokhorov M.E., 2006,
MNRAS, 367, 732

\bibitem{rau2005} Rau A., Kienlin A.V., Hurley K., Lichti G.G., 2005, A\&A, 438, 1175

\bibitem{ruderman2000} Ruderman M.A., Tao L., Kluzniak W., 2000, ApJ, 542, 243

\bibitem{sguera2005} Sguera V. et al., 2005,
A\&A, 444, 221

\bibitem{shakura} Shakura N.I., Sunyaev R.A., 1973, A\&A, 24, 337

\bibitem{smith2006a}  Smith D.M., Heindl W.A., Markwardt C.B., Swank J.H., Negueruela I., Harrison T.E., Huss
L., ApJ, 638, 974

\bibitem{sollerman2005}  Sollerman J., Ostlin G., Fynbo J.P.U., Hjorth J., Fruchter A., Pedersen K., 2005,
New Astron., 11, 103

\bibitem{stern2001} Stern B.E., Tikhomirova Ya., Kompaneets D., Svensson R., Poutanen J., 2001, 563, 80

\bibitem{stern2002} Stern B.E., Tikhomirova Ya., Svensson R., 2002, ApJ, 573,
75

\bibitem{tanvir2005}  Tanvir N.R., Chapman R., Levan A.J., Priddey R.S., 2005,
Nature, 438, 991

\bibitem{thompson1993} Thompson C., Duncan R.C., 1993, ApJ, 408,
194

\bibitem{thompson2001} Thompson T.A., Burrows A., Meyer B.S., 2001, ApJ, 562, 887

\bibitem{tutukov1969} Tutukov A.V., 1969, Nauchnye Informatsii, 11, 69

\bibitem{tutukov2003b} Tutukov A.V., 2003, Astronomy Reports, 47, 637

\bibitem{tutukov2003} Tutukov A.V., Cherepashchuk A.M., 2003, Astronomy Reports, 47, 386

\bibitem{tutukov2004} Tutukov A.V., Pavlyuchenkov Ya.N., 2004, Astronomy Reports, 48, 800

\bibitem{tutukov1979} Tutukov A.V., Yungelson L.R., 1979, Acta Astron., 29, 665

\bibitem{tutukov1993} Tutukov A.V., Yungelson L.R., 1993,
MNRAS, 260, 675

\bibitem{tutukov2002} Tutukov A.V., Yungelson L.R., 2002, Astronomy Reports, 46, 667

\bibitem{tutukov1973a} Tutukov A., Yungelson L., Klayman
A., 1973, Nauchnye Informatsii, v. 27, p.3

\bibitem{umeda2005}  Umeda H., Tominaga N., Maeda K., Nomoto K., 2005, ApJ,
633, L17

\bibitem{usov1975} Usov V.V., Chibisov G.V., 1975, SvA, 19, 115

\bibitem{vanbeveren} Vanbeveren D., de Donger E., van Bever J., van Rensbergen W., de Loore C., 1998, New Astron., 3, 443

\bibitem{heuvel1994} van den Heuvel E.P.J., in Shore S.N., Livio M., van den Heuvel E.P.J., 1994, Interacting Binaries, Springer-Verlag, p. 103

\bibitem{vanlandingham2005} Vanlandingham K. et al., 2005,
AJ, 130, 734

\bibitem{varshavskii} Varshavskii V. I., Tutukov A. V., 1975, SvA, 19, 142

\bibitem{vibe1998} Wiebe D.S., Tutukov A.V., Shustov, B.M., 1998, Astronomy Reports, 42, 1

\bibitem{woods2004a}
Woods P.M., Thompson C., 2004, preprint (astro-ph/0406133)

\bibitem{woosley1993a} Woosley S.E., 1993, ApJ, 405, 273

\end{thebibliography}
\end{document}